\documentstyle[amssymb,12pt,thmsa,sw20lart]{article}
\input{tcilatex}
\begin{document}

\title{Quantum displacements}
\author{Zai-Zhe Zhong \\
%EndAName
Department of Physics, Liaoning Normal University, Dalian 116029, \\
Liaoning, China. E-mail: zhongzaizheh@hotmail.com}
\maketitle

\begin{abstract}
In this paper, first we explain what are the `quantum displacements'. We
establish a group of bases, which contains the coupled bases coupling a
ququart and a bipartite qubit systems. By these bases, we can realize the
quantum displacements. We discuss some possible forms of them. At last, we
point out that a so-call ''non-imprecisely-cloning theorem'' also holds.

PACC numbers: 03.67.Mn, 03.65.Ud, 03.67.Hk.
\end{abstract}

From the original works of BBCJPW[1] and ZZHE[$2$] till now, in the theory
and experiments of modern quantum mechanics, especially in the information
and quantum computer, the task of quantum teleportation and swapping are all
long of the paramount importance. There have been very many related papers
(e.g. see the references in [3,4$]$, and for the multipartite d-level(d$%
\geqslant $3)\ systems, see\ [5-10]). In this paper we shall point out a new
quantum process, the quantum displacements, this process is different from
the all ordinary quantum teleportation and swapping, and more quantum
informations are non-locally transmitted in such processes. Therefore, at
least in the theory, we prove the possibility that in some cases we can use
a new quantum way and classical communications to teleport more informations.

In order to get a clear understanding of basic point in this paper, first we
look upon the simplified figure of general quantum teleportation as follows

\begin{equation}
\begin{array}{lllllll}
\;\;\underline{\text{Alice}} &  & \;\;\text{\underline{Bob}} &  & \;\;%
\underline{\text{Clara}} &  &  \\ 
\mid \left( ?m\right) > & \longleftarrow & \;\mid \left( n\right) ,\left(
m\right) >_\alpha & \longrightarrow \text{ } & \;\;\;\emptyset &  &  \\ 
\;\;\downarrow &  &  &  & \;\;\;\downarrow &  &  \\ 
\mid \left( n\right) ,\left( m\right) >_\beta & \longrightarrow & 
\begin{array}{l}
\text{calssical } \\ 
\text{communications}
\end{array}
& \longrightarrow & \mid \left( m\right) >_\beta & \text{return}%
\longrightarrow & \mid \left( ?m\right) >
\end{array}
\end{equation}
where $\left( n\right) $ and $\left( m\right) ,$ respectively, denote m and
n particles which are in the same kind, Alice holds m particles which are in
an unknown state $\mid \left( ?m\right) >$, Bob (the quantum channel) holds
(n+m) particles, they are in a maximal entangled state $\mid \left( n\right)
,\left( m\right) >_\alpha .$ Bob send $\left( n\right) $ to Alice and $%
\left( m\right) $ to Clara, respectively. Alice make a measurement, then the
state will collapse into a maximal entangled state $\mid \left( n\right)
,\left( m\right) >_\beta $ with equal possibility, simultaneously $\left(
m\right) $ will be in some corresponding state $\mid \left( m\right) >_\beta
.$ When Alice informs her measurement result to Clara by using of the
classical communication, then by using of a determined unitary
transformation $U_{\alpha ,\beta }$, Clara knows that the correct result
should be $U_{\alpha ,\beta }\mid \left( m\right) >_\beta =\mid \left(
?m\right) >$. In this paper, the simplified figure of general case in our
scheme is as 
\begin{equation}
\begin{array}{lllllll}
\;\underline{\text{Alice}} &  & \;\;\;\;\underline{\text{Bob}} &  & \;\;\;\;%
\underline{\text{Clara}} &  &  \\ 
\mid \left( ?m\right) > & \longleftarrow & \;\mid \left( n\right) ,\left(
G\right) >_\alpha & \longrightarrow \text{ } & \;\;\;\;\;\emptyset &  &  \\ 
\;\;\;\downarrow &  &  &  & \;\;\;\;\;\downarrow &  &  \\ 
\mid \left( m\right) ,\left( n\right) >_\beta & \longrightarrow & 
\begin{array}{l}
\text{calssical } \\ 
\text{communications}
\end{array}
& \longrightarrow & \;\mid \left( G\right) >_\beta \longrightarrow & 
\longrightarrow & 
\begin{array}{l}
\text{To know } \\ 
\text{original}\mid \left( ?m\right) >
\end{array}
\\ 
&  &  &  & \;\;\;\;\;\downarrow &  &  \\ 
&  &  &  & 
\begin{array}{l}
\text{To know the } \\ 
\text{change of } \\ 
\text{state of Bob}
\end{array}
&  & 
\end{array}
\end{equation}
where we stress that $\left( G\right) $ denotes G particles, however which
can be in the differential kind from $\left( m\right) $ and $\left( n\right)
,$ the rest symbols denote the similar meaning as in figure (1). When Alice
informs her measurement result to Clara by using of the classical
communications, by using of a determined unitary transformation (see below)
Clara can know what is the original particles $\mid \left( ?m\right) >$ of
Alice , and after this, she also knows that the change of state of Bob (from 
$\mid \left( n\right) ,\left( G\right) >_\alpha $to $\mid \left( m\right)
,\left( n\right) >_\beta ).$ Obviously, our scheme (as in (2)), generally,
is not an ordinary quantum teleportation (the later is a special case of the
when $(G)\equiv \left( m\right) )$. Sum up, the distinctions between schemes
(1) and (2) are:

({\rm i}){\rm \ }In (1), particles in the same kind are transmitted from
Alice to Clara, and by this Clara obtains the information $\mid \left(
?m\right) >,$ but in (2) the particles transmitted change into particles $%
(G) $ which can be in other kind.

({\rm ii}){\rm \ }The informations obtained by Clara in the case (2)
obviously are more than her in the case (1).

({\rm iii}) The type of quantum channel is invariant in (1), but it is
changed in (2).

({\rm iv}){\rm \ }The original particles in (1) and (2) both are broken in
the processes, but the corresponding results are distinct, this means that
schemes (1) and (2) must relate the distinct non-cloning problems,
respectively.

Here we can use a word `displacement' in the chemistry (e.g. in the
generation of hydrogen, $Zn+H_2SO_4\longrightarrow ZnH_2SO_4+H_2\uparrow $,
the states$\mid \left( ?m\right) >,\mid \left( n\right) ,\left( G\right)
>_\alpha ,$ $\mid \left( m\right) ,\left( n\right) >_\beta $ and $\mid
\left( G\right) >$, respectively, correspond to the zinc, the sulphuric
acid, the zinc sulfate, and the hydrogen, etc.), and call the above process
a `quantum displacement'.

In this paper, we only detail the cases of two qubit and a ququart states.
In the first place, we need to establish some quantum channels, a group of
bases\ (the maximal entanglement representations) containing some so-called
`coupled bases'. By these bases, we can realize the quantum displacements,
and discuss their possible forms. In addition, we yet discuss the quantum
displacements in some swapping. At last,we discuss the problem of the
non-cloning theorem, we prove that a so-called `non-imprecisely-cloning
theorem' holds also.

In the following we denote the Hilbert space of quNit states by $H_i^{\left(
N\right) },$ where $i$ is the serial number of the Hilbert space. We shall
consider the following products: the ordinary bipartite ququarts system $H_{%
{\rm I,II}}^{\left( 16\right) }\equiv H_{{\rm I}}^{\left( 4\right) }\otimes
H_{{\rm II}}^{\left( 4\right) }$ , quadripartite qubit system $%
H_{1,2,3,4}^{\left( 16\right) }\equiv H_1^{\left( 2\right) }\otimes
H_2^{\left( 2\right) }\otimes $ $H_3^{\left( 2\right) }\otimes H_4^{\left(
2\right) },$ the homogeneous products $H_{{\rm I,}1,2}^{\left( 16\right)
}\equiv H_{{\rm I}}^{\left( 4\right) }\otimes H_1^{\left( 2\right) }\otimes
H_2^{\left( 2\right) }$ and $H_{1,2,{\rm I}}^{\left( 16\right) }\equiv
H_1^{\left( 2\right) }\otimes H_2^{\left( 2\right) }\otimes H_{{\rm I}%
}^{\left( 4\right) }$. In the first place, we need to point out that
although $H_{{\rm I}}^{\left( 4\right) }$ and $H_1^{\left( 2\right) }\otimes
H_2^{\left( 2\right) }$(or $H_3^{\left( 2\right) }\otimes H_4^{\left(
4\right) }$) both are four-dimensional, they are completely distinct spaces,
e.g. in $H_1^{\left( 2\right) }\otimes H_2^{\left( 2\right) }$ there are
many entangled states, conversely in single $H_{{\rm I}}^{\left( 4\right) }$
we generally don't consider the entanglement, etc. Thus, $H_{{\rm I,II}%
}^{\left( 16\right) },H_{1,2,3,4}^{\left( 16\right) }$ and $H_{1,2,{\rm I}%
}^{\left( 16\right) }$ (or $H_{1,2,{\rm I}}^{\left( 16\right) }$) are the
distinct spaces. However, they all are 16-dimensional Hilbert spaces and
have some similar constructions, this point is quite important in this paper
(Recently the quantum teleportation problem of $H_{1,2,3,4}^{\left(
16\right) }$ has been discussed in [11]). In the following, in the natural
bases we always write $\mid i>,i=0,1,2,3$ for the ququarts, and $\mid
i>=\mid rs>$ for the bipartite qubit, where $rs$ $=00,01,10,11$.

Now, we take formally the basis $\left\{ \mid W_\alpha >,\mid X_\alpha
>,\mid Y_\alpha >,\mid Z_\alpha >\right\} $ in an uniform as 
\begin{eqnarray}
&\mid &W_\alpha >=\frac 12\left( \mid A_\alpha >+\mid B_\alpha >+\mid
C_\alpha >+\mid D_\alpha >\right)  \nonumber \\
&\mid &X_i>=\frac 12\left( \mid A_\alpha >+\mid B_\alpha >-\mid C_\alpha
>-\mid D_\alpha >\right)  \nonumber \\
&\mid &Y_i>=\frac 12\left( \mid A_\alpha >-\mid B_\alpha >+\mid C_\alpha
>-\mid D_\alpha >\right) \\
&\mid &Z_i>=\frac 12\left( \mid A_\alpha >-\mid B_\alpha >-\mid C_\alpha
>+\mid D_\alpha >\right)  \nonumber
\end{eqnarray}
If in Eq.(3) we substitute the natural bases ($\alpha =i=0,1,2,3$) 
\begin{eqnarray}
&\mid &A_i>=\mid i_{{\rm I}}>\mid 0_{{\rm II}}>,\mid B_i>=\mid \left( i+1%
\func{mod}4\right) _{{\rm I}}>\mid 1_{{\rm II}}>  \nonumber \\
&\mid &C_i>=\mid \left( i+2\func{mod}4\right) _{{\rm I}}>\mid 2_{{\rm II}%
}>,\mid D_i>=\mid \left( i+3\func{mod}4\right) _{{\rm I}}>\mid 3_{{\rm II}}>
\end{eqnarray}
then we obtain a complete orthogonal basis $\left\{ \mid W_i^{\left( {\rm %
I,II}\right) }>,\mid X_i^{\left( {\rm I,II}\right) }>,\mid Y_i^{\left( {\rm %
I,II}\right) }>,\mid Z_i^{\left( {\rm I,II}\right) }>\right\} $ of $H_{{\rm %
I,II}}^{\left( 16\right) }$. If in Eq.(3) we substitute the natural bases ($%
\alpha =rs=00,01,10,11)$ 
\begin{eqnarray}
&\mid &A_{rs}>=\mid r_1s_20_30_4>,\mid B_{rs}>=\mid \left( 1-r\right)
_1\left( 1-s\right) _20_31_4>  \nonumber \\
&\mid &C_{rs}>=\mid \left( 1-r\right) _1s_21_30_4>,\mid D_{rs}>=\mid
r_1\left( 1-s\right) _21_31_4>
\end{eqnarray}
then we obtain a complete orthogonal basis $\left\{ \mid W_{rs}^{\left(
1234\right) }>,\mid X_{rs}^{\left( 1234\right) }>,\mid Y_{rs}^{\left(
1234\right) }>,\mid Z_{rs}^{\left( 1234\right) }>\right\} $ of $%
H_{1,2,3,4}^{\left( 16\right) }$. We are especially interesting to the
coupled bases, i.e. in Eq.(3) we substitute 
\begin{eqnarray}
&\mid &A_{rs}>=\mid 0_{{\rm I}}>\mid r_1s_2>,\mid B_{rs}>=\mid 1_{{\rm I}%
}>\mid \left( 1-r\right) _1\left( 1-s\right) _2>  \nonumber \\
&\mid &C_{rs}>=\mid 2_{{\rm I}}>\mid \left( 1-r\right) _1s_2>,\mid
D_{rs}>=\mid 3_{{\rm I}}>\mid r_1\left( 1-s\right) _2>
\end{eqnarray}
then we obtain a basis $\left\{ \mid W_{rs}^{\left( {\rm I,}12\right)
}>,\mid X_{rs}^{\left( {\rm I,}12\right) }>,\mid Y_{rs}^{\left( {\rm I,}%
12\right) }>,\mid Z_{rs}^{\left( {\rm I,}12\right) }>\right\} $ of the
Hilbert space $H_{{\rm I,}12}^{\left( 16\right) }$, which is a coupled basis
coupling a ququart system and a bipartite qubit system$.$ Completely
similarly, $\left\{ \mid W_{rs}^{\left( 12,{\rm I}\right) }>,\mid
X_{rs}^{\left( 12,{\rm I}\right) }>,\mid Y_{rs}^{\left( 12,{\rm I}\right)
}>,\mid Z_{rs}^{\left( 12,{\rm I}\right) }>\right\} $ for $H_{12,{\rm I}%
}^{\left( 16\right) }$. Here we must stress that for all the above bases the
transformation relations, from $\left\{ \mid W_\alpha >,\mid X_\alpha >,\mid
Y_\alpha >,\mid Z_\alpha >\right\} $ to $\left\{ \mid A_\alpha >,\mid
B_\alpha >,\mid C_\alpha >,\mid D_\alpha >\right\} $, where $\alpha =i$ or $%
\alpha =rs$, are the same form (this point is important for the purpose of
this paper), i.e. 
\begin{eqnarray}
&\mid &A_\alpha >=\frac 12\left( \mid W_\alpha >+\mid X_\alpha >+\mid
Y_\alpha >+\mid Z_\alpha >\right)  \nonumber \\
&\mid &B_\alpha >=\frac 12\left( \mid W_\alpha >+\mid X_\alpha >-\mid
Y_\alpha >-\mid Z_\alpha >\right)  \nonumber \\
&\mid &C_\alpha >=\frac 12\left( \mid W_\alpha >-\mid X_\alpha >+\mid
Y_\alpha >-\mid Z_\alpha >\right) \\
&\mid &D_\alpha >=\frac 12\left( \mid W_\alpha >-\mid X_\alpha >-\mid
Y_\alpha >+\mid Z_\alpha >\right)  \nonumber
\end{eqnarray}
The above bases, in fact, give some maximal entanglement representations of $%
H_{{\rm I,II}}^{\left( 16\right) },H_{1,2,3,4}^{\left( 16\right) }$ and $%
H_{1,2,{\rm I}}^{\left( 16\right) }$ (or $H_{1,2,{\rm I}}^{\left( 16\right)
} $) (however we need not to discuss it in this paper).

Now we suppose that Alice holds the particle {\rm I} which is in an unknown
four-level pure-state $\mid \phi ^{\left( {\rm I}\right) }>=\alpha \mid 0_{%
{\rm I}}>+\beta \mid 1_{{\rm I}}>+\gamma \mid 2_{{\rm I}}>+\delta \mid 3_{%
{\rm I}}>,$ Clara is in the remote places from Alice. Bob holds two
particles {\rm II} (four-level) and 1, 2 ( two-level state) and she makes
them to be in a basic state, for instance, in $\mid X_1^{\left( {\rm II,}%
12\right) }>=\frac 12\left( \mid 1_{{\rm II}}>\mid 0_10_2>+\mid 2_{{\rm II}%
}>\mid 0_11_2>-\mid 3_{{\rm II}}>\mid 1_10_2>-\mid 0_{{\rm II}}>\mid
1_11_2>\right) $, then the total state is $\mid \Psi _{total}>=\mid \phi
^{\left( {\rm I}\right) }>\mid X_1^{\left( {\rm II,}12\right) }>=\frac
12\left( 
\begin{array}{c}
\mid 0_{{\rm I}}>\mid 1_{{\rm II}}>\alpha \mid 0_10_2> \\ 
+\cdots -\mid 3_{{\rm I}}>\mid 0_{{\rm II}}>\delta \mid 1_11_2>
\end{array}
\right) $. According to Eqs.(4) and (7), every $\mid i_{{\rm I}}>\mid j_{%
{\rm II}}>$ always can be expressed by $\mid W_k^{\left( {\rm I,II}\right)
}>,\;\mid X_k^{\left( {\rm I,II}\right) }>,\;$ $\mid Y_k^{\left( {\rm I,II}%
\right) }>$ and $\mid Z_k^{\left( {\rm I,II}\right) }>.$ Substitute them and
reorganize, the last result is 
\begin{eqnarray}
&\mid &\Psi _{total}>=\sum_{i=0}^3\left( 
\begin{array}{c}
\mid W_i^{\left( {\rm I,II}\right) }>\mid \phi _{W_i}^{\left( 12\right)
}>+\mid X_i^{\left( {\rm I,II}\right) }>\mid \phi _{X_i}^{\left( 12\right) }>
\\ 
+\mid Y_i^{\left( {\rm I,II}\right) }\mid \phi _{Y_i}^{\left( 12\right)
}>+\mid Z_i^{\left( {\rm I,II}\right) }>\mid \phi _{Z_i}^{\left( 12\right) }>
\end{array}
\right)  \nonumber \\
&\equiv &\frac 14\left\{ 
\begin{array}{c}
\mid W_0^{\left( {\rm I,II}\right) }>U_{W_0}^{\dagger }+\mid W_1^{\left( 
{\rm I,II}\right) }>U_{W_1}^{\dagger }+\mid W_2^{\left( {\rm I,II}\right)
}>U_{W_2}^{\dagger }+\mid W_3^{\left( {\rm I,II}\right) }>U_{W_3}^{\dagger
}\ \, \\ 
+\mid X_0^{\left( {\rm I,II}\right) }>U_{X_0}^{\dagger }+\mid X_1^{\left( 
{\rm I,II}\right) }>U_{X_1}^{\dagger }+\mid X_2^{\left( {\rm I,II}\right)
}>U_{X2}^{\dagger }+\mid X_3^{\left( {\rm I,II}\right) }>U_{X_3}^{\dagger }\
\, \\ 
+\mid Y_0^{\left( {\rm I,II}\right) }>U_{Y_0}^{\dagger }+\mid Y_1^{\left( 
{\rm I,II}\right) }>U_{Y_1}^{\dagger }+\mid Y_2^{\left( {\rm I,II}\right)
}>U_{Y_2}^{\dagger }+\mid Y_3^{\left( {\rm I,II}\right) }>U_{Y3}^{\dagger }\
\, \\ 
+\mid Z_0^{\left( {\rm I,II}\right) }>U_{Z_0}^{\dagger }+\mid Z_1^{\left( 
{\rm I,II}\right) }>U_{Z_1}^{\dagger }+\mid Z_2^{\left( {\rm I,II}\right)
}>U_{Z_2}^{\dagger }+\mid Z_3^{\left( {\rm I,II}\right) }>U_{Z_3}^{\dagger }
\end{array}
\right\}  \nonumber \\
&&\times \left( \alpha \mid 0_10_2>+\beta \mid 0_11_2>+\gamma \mid
1_10_2>+\delta \mid 1_11_2>\right)
\end{eqnarray}
where all $U_{\bullet }$ $\left( \bullet =W_0,W_1,\cdots ,Z_3\right) $ are
unitary matrixes, 
\begin{eqnarray}
U_{W_0} &=&\left[ 
\begin{array}{llll}
0 & 0 & 0 & -1 \\ 
1 & 0 & 0 & 0 \\ 
0 & 1 & 0 & 0 \\ 
0 & 0 & -1 & 0
\end{array}
\right] ,\;U_{W_1}=\left[ 
\begin{array}{llll}
0 & 0 & -1 & 0 \\ 
0 & 0 & 0 & -1 \\ 
1 & 0 & 0 & 0 \\ 
0 & 1 & 0 & 0
\end{array}
\right]  \nonumber \\
\;U_{W_2} &=&\left[ 
\begin{array}{llll}
0 & 1 & 0 & 0 \\ 
0 & 0 & -1 & 0 \\ 
0 & 0 & 0 & -1 \\ 
1 & 0 & 0 & 0
\end{array}
\right] ,\;U_{W_3}=\left[ 
\begin{array}{llll}
1 & 0 & 0 & 0 \\ 
0 & 1 & 0 & 0 \\ 
0 & 0 & -1 & 0 \\ 
0 & 0 & 0 & -1
\end{array}
\right]  \nonumber \\
U_{X_0} &=&\left[ 
\begin{array}{llll}
0 & 0 & 0 & -1 \\ 
1 & 0 & 0 & 0 \\ 
0 & -1 & 0 & 0 \\ 
0 & 0 & -1 & 0
\end{array}
\right] ,\;U_{X_1}=\left[ 
\begin{array}{llll}
0 & 0 & 1 & 0 \\ 
0 & 0 & 0 & -1 \\ 
1 & 0 & 0 & 0 \\ 
0 & -1 & 0 & 0
\end{array}
\right]  \nonumber \\
\;U_{X_2} &=&\left[ 
\begin{array}{llll}
0 & -1 & 0 & 0 \\ 
0 & 0 & 1 & 0 \\ 
0 & 0 & 0 & -1 \\ 
1 & 0 & 0 & 0
\end{array}
\right] ,\;U_{X_3}=\left[ 
\begin{array}{llll}
-1 & 0 & 0 & 0 \\ 
0 & -1 & 0 & 0 \\ 
0 & 0 & -1 & 0 \\ 
0 & 0 & 0 & -1
\end{array}
\right]  \nonumber \\
U_{Y_0} &=&\left[ 
\begin{array}{llll}
0 & 0 & 0 & -1 \\ 
-1 & 0 & 0 & 0 \\ 
0 & 1 & 0 & 0 \\ 
0 & 0 & 1 & 0
\end{array}
\right] ,\;U_{Y_1}=\left[ 
\begin{array}{llll}
0 & 0 & 1 & 0 \\ 
0 & 0 & 0 & -1 \\ 
-1 & 0 & 0 & 0 \\ 
0 & 1 & 0 & 0
\end{array}
\right] \\
\;U_{Y_2} &=&\left[ 
\begin{array}{llll}
0 & -1 & 0 & 0 \\ 
0 & 0 & -1 & 0 \\ 
0 & 0 & 0 & -1 \\ 
-1 & 0 & 0 & 0
\end{array}
\right] ,\;U_{Y_3}=\left[ 
\begin{array}{llll}
-1 & 0 & 0 & 0 \\ 
0 & 1 & 0 & 0 \\ 
0 & 0 & 1 & 0 \\ 
0 & 0 & 0 & -1
\end{array}
\right]  \nonumber \\
U_{Z_0} &=&\left[ 
\begin{array}{llll}
0 & 0 & 0 & -1 \\ 
-1 & 0 & 0 & 0 \\ 
0 & -1 & 0 & 0 \\ 
0 & 0 & -1 & 0
\end{array}
\right] ,\;U_{Z_1}=\left[ 
\begin{array}{llll}
0 & 0 & -1 & 0 \\ 
0 & 0 & 0 & -1 \\ 
-1 & 0 & 0 & 0 \\ 
0 & -1 & 0 & 0
\end{array}
\right]  \nonumber \\
\;U_{Z_2} &=&\left[ 
\begin{array}{llll}
0 & 1 & 0 & 0 \\ 
0 & 0 & 1 & 0 \\ 
0 & 0 & 0 & -1 \\ 
-1 & 0 & 0 & 0
\end{array}
\right] ,\;U_{Z_3}=\left[ 
\begin{array}{llll}
1 & 0 & 0 & 0 \\ 
0 & -1 & 0 & 0 \\ 
0 & 0 & -1 & 0 \\ 
0 & 0 & 0 & -1
\end{array}
\right]  \nonumber
\end{eqnarray}

This means that when Bob sends the particles {\rm II} to Alice, and sends
the particles $1,2$ to Clara, and Alice makes a associated measurement of
particles {\rm I }and{\rm \ II} , then she will obtain one and only one of
16 basic states $\left\{ \mid W_i^{\left( {\rm I,II}\right) }>,\mid
X_i^{\left( {\rm I,II}\right) }>,\mid Y_i^{\left( {\rm I,II}\right) }>,\mid
Z_i^{\left( {\rm I,II}\right) }>\right\} \left( i=0,1,2,3\right) $ with the
probability $\frac 1{16}$ (assume that there are such instruments$)$.
Simultaneously the particle $1$ and $2$ must be in a corresponding one state
of $\left\{ \mid \phi _{W_i}^{\left( 12\right) }>,\mid \phi _{X_i}^{\left(
12\right) }>,\mid \phi _{Y_i}^{\left( 12\right) }>,\mid \phi _{Z_i}^{\left(
12\right) }\right\} .$ When Alice informs Clara of her result $\mid \mu $ 
\TEXTsymbol{>} ($\mid \mu $ \TEXTsymbol{>} is one and only one of $\mid
W_i>,\mid X_i>,\mid Y_i>,\mid Z_i>$ with probability $\frac 1{16}$) by some
classical communications, then Clara at once knows the correct result should
be $\mid \phi ^3>=U_{\mu _i}\mid \phi _{\mu _i}^3>$. In addition, after this
Clara also knows the change of state of Bob is from $\mid X_1^{\left( {\rm %
II,}12\right) }>$ to $\mid \mu >.$ Similarly, we yet use other basis
vectors, e.g. $\mid X_2>$ $\mid Y_1>,\mid Y_2>,\cdots ,$ etc. Now, the
quantum displacements are completed. Here we notice that the particle
`inputted' (Alice ) is one, but the particles `outputted' (Clara) are two,
and if Clara wants to know what is the original (four-level) particle, then
they must wait for Alice (notice that $\mid \phi ^{\left( {\rm I}\right) }>$
is unknown for Alice) to inform to them of her measurement result by some
classical communications.

The above process (the related calculations as in Eqs. (8) and (9) have been
omitted) can be simply figured as (it is like somewhat a chemical equation) 
\begin{equation}
\left( {\rm i}\right) \;\bullet _{{\rm I}}^{\left( 4\right) }+\bullet _{{\rm %
II}}^{\left( 4\right) }**\bullet _1^{\left( 2\right) }**\bullet _2^{\left(
2\right) }\longrightarrow \bullet _{{\rm I}}^{\left( 4\right) }**\bullet _{%
{\rm II}}^{\left( 4\right) }+\bullet _1^{\left( 2\right) }**\bullet
_2^{\left( 2\right) }
\end{equation}
where symbol $**$ denotes some possible entanglement. Similarly, by using of
the similar ways$,$ we can obtain the following results, of which the
calculations are completely similar, 
\begin{eqnarray}
\left( {\rm ii}\right) \;\bullet _{{\rm I}}^{\left( 4\right) }+\bullet
_1^{\left( 2\right) }**\bullet _2^{\left( 2\right) }**\bullet _{{\rm II}%
}^{\left( 4\right) } &\longrightarrow &\bullet _1^{\left( 2\right)
}**\bullet _2^{\left( 2\right) }**\bullet _{{\rm I}}^{\left( 4\right)
}+\bullet _{{\rm II}}^{\left( 4\right) }  \nonumber \\
\left( {\rm iii}\right) \;\bullet _{{\rm I}}^{\left( 4\right) }+\bullet _{%
{\rm II}}^{\left( 4\right) }**\bullet _{{\rm III}}^{\left( 4\right) }
&\longrightarrow &\bullet _{{\rm I}}^{\left( 4\right) }**\bullet _{{\rm II}%
}^{\left( 4\right) }+\bullet _{{\rm III}}^{\left( 4\right) }  \nonumber \\
\left( {\rm iv}\right) \;\bullet _{{\rm I}}^{\left( 4\right) }+\bullet _{%
{\rm II}}^{\left( 4\right) }**\bullet _1^{\left( 2\right) }**\bullet
_2^{\left( 2\right) } &\longrightarrow &\bullet _{{\rm I}}^{\left( 4\right)
}**\bullet _{{\rm II}}^{\left( 4\right) }+\bullet _1^{\left( 2\right)
}**\bullet _2^{\left( 2\right) }  \nonumber \\
\left( {\rm v}\right) \;\bullet _1^{\left( 2\right) }**\bullet _2^{\left(
2\right) }+\bullet _{{\rm I}}^{\left( 4\right) }**\bullet _3^{\left(
2\right) }**\bullet _4^{\left( 2\right) } &\longrightarrow &\bullet
_1^{\left( 2\right) }**\bullet _2^{\left( 2\right) }**\bullet _{{\rm I}%
}^{\left( 4\right) }+\bullet _3^{\left( 2\right) }**\bullet _4^{\left(
2\right) } \\
\left( {\rm vi}\right) \;\bullet _1^{\left( 2\right) }**\bullet _2^{\left(
2\right) }+\bullet _{{\rm II}}^{\left( 4\right) }**\bullet _3^{\left(
2\right) }**\bullet _4^{\left( 2\right) } &\longrightarrow &\bullet
_1^{\left( 2\right) }**\bullet _2^{\left( 2\right) }**\bullet _{{\rm II}%
}^{\left( 4\right) }+\bullet _3^{\left( 2\right) }**\bullet _4^{\left(
2\right) }  \nonumber \\
\left( {\rm vii}\right) \;\bullet _1^{\left( 2\right) }**\bullet _2^{\left(
2\right) }+\bullet _3^{\left( 2\right) }**\bullet _4^{\left( 2\right)
}**\bullet _{{\rm II}}^{\left( 4\right) } &\longrightarrow &\bullet
_1^{\left( 2\right) }**\bullet _2^{\left( 2\right) }**\bullet _3^{\left(
2\right) }**\bullet _4^{\left( 2\right) }+\bullet _{{\rm II}}^{\left(
4\right) }  \nonumber \\
\left( {\rm viii}\right) \;\bullet _1^{\left( 2\right) }**\bullet _2^{\left(
2\right) }+\bullet _3^{\left( 2\right) }**\bullet _4^{\left( 2\right)
}**\bullet _5^{\left( 2\right) }**\bullet _6^{\left( 2\right) }
&\longrightarrow &\bullet _1^{\left( 2\right) }**\bullet _2^{\left( 2\right)
}**\bullet _3^{\left( 2\right) }**\bullet _4^{\left( 2\right) }+\bullet
_5^{\left( 2\right) }**\bullet _6^{\left( 2\right) }  \nonumber
\end{eqnarray}
Obviously, the cases of $\left( {\rm ii}\right) ,\left( {\rm iii}\right) $
and $\left( {\rm viii}\right) ,$ in fact, are the ordinary quantum
teleportation, $\left( {\rm viii}\right) $ has been considered in [11].

In the quantum swapping there may be yet the displacements. For instance, we
suppose that Alice holds the particle {\rm I}, Bob holds the particles 1, 2
, 3, 4 and Clara holds the particle {\rm II}. The particles {\rm I }and{\rm %
\ }1, 2 are in the entangled state $\mid X_1^{\left( {\rm I,}12\right) }>$,
and the particles 3, 4 and {\rm II }are in the entangled state $\mid
X_1^{\left( 34,{\rm II}\right) }>$. Therefore the total state is $\mid \Phi
_{total}>=\mid X_1^{\left( {\rm I,}12\right) }>\mid X_1^{\left( 34,{\rm II}%
\right) }>.$ We can make the following direct calculation: 
\begin{eqnarray}
&\mid &\Phi _{total}>=\frac 14\left( \mid 1_{{\rm I}}>\mid 0_10_2>+\mid 2_{%
{\rm I}}>\mid 0_11_2>-\mid 3_{{\rm I}}>\mid 1_10_2>-\mid 0_{{\rm I}}>\mid
1_11_2>\right)  \nonumber \\
&&\otimes \left( \mid 0_31_4>\mid 0_{{\rm II}}>+\mid 1_30_4>\mid 1_{{\rm II}%
}>-\mid 1_31_4>\mid 2_{{\rm II}}>-\mid 0_30_4>\mid 3_{{\rm II}}>\right) 
\nonumber \\
&=&\frac 14\left( \mid 1_{{\rm I}}>\mid 0_10_2>\mid 0_31_4>\mid 0_{{\rm II}%
}>\cdots +\mid 0_{{\rm I}}>\mid 1_11_2>\mid 0_30_4>\mid 3_{{\rm II}}>\right)
\\
&=&\frac 18\left\{ 
\begin{array}{c}
\mid 1_{{\rm I}}>\left( \mid W_3^{\left( 1234\right) }>-\mid X_3^{\left(
1234\right) }>-\mid Y_3^{\left( 1234\right) }>+\mid Z_3^{\left( 1234\right)
}>\right) \mid 0_{{\rm I}}> \\ 
\cdots +\mid 0_{{\rm I}}>\left( \mid W_3^{\left( 1234\right) }>+\mid
X_3^{\left( 1234\right) }>+\mid Y_3^{\left( 1234\right) }>+\mid Z_3^{\left(
1234\right) }>\right) \mid 3_{{\rm I}}>
\end{array}
\right\}  \nonumber
\end{eqnarray}
For $\mid 1_1>\mid 0_4>$, $\mid 1_1>\mid 1_4>,\cdots ,\mid 0_1>\mid 3_4>$ we
use Eq.($7$), and in $H_{1,2,3,4}^{\left( 16\right) }\otimes H_{{\rm I,II}%
}^{\left( 16\right) }$ rewrite $\mid \Phi _{total}>$, we find, in fact, 
\begin{equation}
\mid \Phi _{total}>=\frac 14\left\{ 
\begin{array}{c}
\mid W_0^{\left( 1234\right) }>\mid Z_2^{\left( {\rm I,II}\right) }>-\mid
W_1^{\left( 1234\right) }>\mid X_3^{\left( {\rm I,II}\right) }> \\ 
-\mid W_2^{\left( 1234\right) }>\mid Y_0^{\left( {\rm I,II}\right) }>+\mid
W_3^{\left( 1234\right) }>\mid W_1^{\left( {\rm I,II}\right) }> \\ 
+\mid X_0^{\left( 1234\right) }>\mid X_2^{\left( {\rm I,II}\right) }>-\mid
X_1^{\left( 1234\right) }>\mid X_3^{\left( {\rm I,II}\right) }> \\ 
-\mid X_2^{\left( 1234\right) }>\mid X_0^{\left( {\rm I,II}\right) }>-\mid
X_3^{\left( 1234\right) }>\mid X_1^{\left( {\rm I,II}\right) }> \\ 
-\mid Y_0^{\left( 1234\right) }>\mid W_2^{\left( {\rm I,II}\right) }>+\mid
Y_1^{\left( 1234\right) }>\mid Y_3^{\left( {\rm I,II}\right) }> \\ 
+\mid Y_2^{\left( 1234\right) }>\mid Z_0^{\left( {\rm I,II}\right) }>-\mid
Y_3^{\left( 1234\right) }>\mid Z_1^{\left( {\rm I,II}\right) }> \\ 
-\mid Z_0^{\left( 1234\right) }>\mid Y_2^{\left( {\rm I,II}\right) }>-\mid
Z_1^{\left( 1234\right) }>\mid Z_3^{\left( {\rm I,II}\right) }> \\ 
+\mid Z_2^{\left( 1234\right) }>\mid W_0^{\left( {\rm I,II}\right) }>+\mid
Z_3^{\left( 1234\right) }>\mid Z_1^{\left( {\rm I,II}\right) }>
\end{array}
\right\}
\end{equation}
This means that when Bob makes an associated measurement of particles 1, 2,
3, 4, then the wave function $\mid \Phi _{total}>$ will collapse to only one
of the above 16 states (say, $\mid W_1^{\left( 1234\right) }>$ ) with
probability $\frac 1{16}$, then there appear one corresponding entanglement
(say, $\mid X_3^{\left( {\rm I,II}\right) }>)$ between particles {\rm I} and 
{\rm II}, etc. The above process can be transposed, i.e. the entanglement
among particles 1, 2, 3, 4 and the entanglement between particles {\rm I}
and {\rm II }are swapped to the entanglement among 1, 2, {\rm I }and the
entanglement among particles 3, 4, {\rm II. }The above two processes can be%
{\rm \ }figured simply by 
\begin{equation}
\left( {\rm i}\right) \; 
\begin{array}{lllllll}
\;\;\;\;\;\;\bullet _{{\rm I}}^{\left( 4\right) } &  & \;\;\;\;\;\bullet _{%
{\rm II}}^{\left( 4\right) } &  & \bullet _{{\rm I}}^{\left( 4\right) } &  & 
\bullet _1^{\left( 2\right) }**\bullet _2^{\left( 2\right) } \\ 
\;\;\;\;\;\; 
\begin{array}{l}
\ast \\ 
\ast
\end{array}
& + & \;\;\;\;\;\; 
\begin{array}{l}
\ast \\ 
\ast
\end{array}
& \rightleftarrows & \; 
\begin{array}{l}
\ast \\ 
\ast
\end{array}
& + & \;\;\;\;\;\; 
\begin{array}{l}
\ast \\ 
\ast
\end{array}
\\ 
\bullet _1^{\left( 2\right) }**\bullet _2^{\left( 2\right) } &  & \bullet
_3^{\left( 2\right) }**\bullet _4^{\left( 2\right) } &  & \bullet _{{\rm II}%
}^{\left( 4\right) } &  & \bullet _3^{\left( 2\right) }**\bullet _4^{\left(
2\right) }
\end{array}
\end{equation}
where $\longleftarrow $ denotes the inversion of $\longrightarrow .$
Similarly, for other cases ($\mid X_2>,\mid Y_1>$, $\cdots ,$ etc.).
Similarly, we can yet write the following swapping, of which the related
calculations are completely similar, 
\begin{eqnarray}
\left( {\rm ii}\right) \; && 
\begin{array}{lllllll}
\;\;\;\;\;\;\bullet _{{\rm I}}^{\left( 4\right) } &  & \bullet _5^{\left(
2\right) }**\bullet _6^{\left( 2\right) } &  & \;\;\;\;\;\;\bullet _{{\rm I}%
}^{\left( 4\right) } &  & \bullet _1^{\left( 2\right) }**\bullet _2^{\left(
2\right) } \\ 
\;\;\;\;\;\; 
\begin{array}{l}
\ast \\ 
\ast
\end{array}
& + & \;\;\;\;\;\; 
\begin{array}{l}
\ast \\ 
\ast
\end{array}
& \rightleftarrows & \;\;\;\;\;\; 
\begin{array}{l}
\ast \\ 
\ast
\end{array}
& + & \;\;\;\;\;\; 
\begin{array}{l}
\ast \\ 
\ast
\end{array}
\\ 
\bullet _1^{\left( 2\right) }**\bullet _2^{\left( 2\right) } &  & \bullet
_3^{\left( 2\right) }**\bullet _4^{\left( 2\right) } &  & \bullet _5^{\left(
2\right) }**\bullet _6^{\left( 2\right) } &  & \bullet _3^{\left( 2\right)
}**\bullet _4^{\left( 2\right) }
\end{array}
\nonumber \\
&&\text{ }  \nonumber \\
\left( {\rm iii}\right) \; && 
\begin{array}{lllllll}
\;\;\;\;\;\;\;\bullet _{{\rm I}}^{\left( 4\right) } &  & \bullet _3^{\left(
2\right) }**\bullet _4^{\left( 2\right) } &  & \;\;\;\;\;\;\;\bullet _{{\rm I%
}}^{\left( 4\right) } &  & \bullet _1^{\left( 2\right) }**\bullet _2^{\left(
2\right) } \\ 
\;\;\;\;\;\; 
\begin{array}{l}
\ast \\ 
\ast
\end{array}
& + & \;\;\;\;\;\; 
\begin{array}{l}
\ast \\ 
\ast
\end{array}
& \rightleftarrows & \;\;\;\;\;\; 
\begin{array}{l}
\ast \\ 
\ast
\end{array}
& + & \;\;\;\;\;\; 
\begin{array}{l}
\ast \\ 
\ast
\end{array}
\\ 
\bullet _1^{\left( 2\right) }**\bullet _2^{\left( 2\right) } &  & 
\;\;\;\;\;\;\;\bullet _{{\rm II}}^{\left( 4\right) } &  & \bullet _3^{\left(
2\right) }**\bullet _4^{\left( 2\right) } &  & \;\;\;\;\;\;\;\bullet _{{\rm %
II}}^{\left( 4\right) }
\end{array}
\\
&&\text{\/}  \nonumber \\
\left( {\rm iv}\right) \; && 
\begin{array}{lllllll}
\;\;\;\;\;\;\;\bullet _{{\rm I}}^{\left( 4\right) } &  & \;\bullet _{{\rm III%
}}^{\left( 4\right) } &  & \bullet _{{\rm I}}^{\left( 4\right) } &  & 
\;\;\;\;\;\;\bullet _{{\rm II}}^{\left( 4\right) } \\ 
\;\;\;\;\;\;\; 
\begin{array}{l}
\ast \\ 
\ast
\end{array}
& + & \; 
\begin{array}{l}
\ast \\ 
\ast
\end{array}
& \rightleftarrows & 
\begin{array}{l}
\ast \\ 
\ast
\end{array}
& + & \;\;\;\;\;\; 
\begin{array}{l}
\ast \\ 
\ast
\end{array}
\\ 
\bullet _1^{\left( 2\right) }**\bullet _2^{\left( 2\right) } &  & \;\bullet
_{{\rm II}}^{\left( 4\right) } &  & \bullet _{{\rm III}}^{\left( 4\right) }
&  & \bullet _1^{\left( 2\right) }**\bullet _2^{\left( 2\right) }
\end{array}
\nonumber
\end{eqnarray}
etc.

At last, since, generally, the quantum teleportation should reflect the
effect of the non-cloning theorem[12], here we also mention the problem of
the non-cloning theorem. According the non-cloning theorem, in the quantum
mechanics the following precise cloning process 
\begin{equation}
\mid Q>\mid \Psi ^{\left( 2\right) }>\longrightarrow \mid Q_{\Psi ^{\left(
2\right) }}>\mid \Psi ^{\left( 2\right) }>\mid \Psi ^{\left( 2\right) }>
\end{equation}
is impossible, where $\mid \Psi ^{\left( 2\right) }>=\alpha \mid 0>+\beta
\mid 1>$ is an arbitrary unknown qubit state, $\mid Q>$ and $\mid Q_{\Psi
^{\left( 2\right) }}>$ are the states of cloning machines. Of course, the
following precise cloning process 
\begin{equation}
\mid Q>\mid \Psi ^{\left( 4\right) }>\longrightarrow \mid Q_{\Psi ^{\left(
4\right) }}>\mid \Psi ^{\left( 4\right) }>\mid \Psi ^{\left( 4\right) }>
\end{equation}
is impossible, where $\mid \Psi ^{\left( 4\right) }>=\alpha \mid 0>+\beta
\mid 1>+\gamma \mid 2>+\delta \mid 3>$ is an arbitrary unknown ququart
state. However, from the above discussions in this paper, we see that the
roles of the ququart state $\mid \Psi ^{\left( 4\right) }>=\alpha \mid
0>+\beta \mid 1>+\gamma \mid 2>+\delta \mid 3>$ and of the bipartite qubit
state $\mid \Psi ^{\left( 2,2\right) }>=\alpha \mid 00>+\beta \mid
01>+\gamma \mid 10>+\delta \mid 11>$, in fact, are very similar in many
cases, this means that $\mid \Psi ^{\left( 2,2\right) }>$ can be regarded as
some ''non-imprecise'' copy of $\mid \Psi ^{\left( 4\right) }>,$ or
conversely. Then there appear a problem that are the following ''imprecise''
cloning processes 
\begin{eqnarray}
&\mid &Q>\mid \Psi ^{\left( 2,2\right) }>\longrightarrow \mid Q_{\Psi
^{\left( 2,2\right) }}>\mid \Psi ^{\left( 2,2\right) }>\mid \Psi ^{\left(
4\right) }>  \nonumber \\
&\mid &Q>\mid \Psi ^{\left( 4\right) }>\longrightarrow \mid Q_{\Psi ^{\left(
4\right) }}>\mid \Psi ^{\left( 4\right) }>\mid \Psi ^{\left( 2,2\right) }>
\end{eqnarray}
allowable? We easily prove that (it is completely similar to the proof of
non-cloning theorem[12]) the above `imprecise cloning' processes still are
impossible, i.e. such a `non-imprecisely-cloning theorem' holds also. In the
above processes of quantum displacements, the original particle or pair of
particles of Alice must be broken, this facts just reflect the correctness
of the non-imprecisely-cloning theorem, and reflect yet the profound meaning
of the non-cloning theorem.

{\bf Discussion.}{\it \ }Establish some similar bases, the method in this
paper can be generalized to some higher dimensional cases, and we can yet
obtain some corresponding results, which will be discussed elsewhere.

{\bf Conclusion:}{\it \ }There are new information processes, the quantum
displacements. For some 16-dimensional quantum systems we can construct a
group of complete orthogonal bases, especially the coupled bases. By using
of these bases, the quantum displacements can be realized. In addition, a
so-called `non-imprecisely-cloning theorem' holds also, and its effects are
reflected in the quantum displacements.


\begin{thebibliography}{99}
\bibitem{}  C. H. Bennett, G. Brassard, C. Cr\'{e}peau, R. Jozsa, A. Peres,
and W. K. Wootters, Phys. Rev. Lett., {\bf 70}(1993)1895.

\bibitem{}  M. Zukowski, A. Zeilinger, M. A. Horne, and A. Ekert, Phys. Rev.
Lett., {\bf 71}(1993)4287.

\bibitem{}  M. A. Neilsen and I. L. Chuang, Quantum Computation and Quantum
Information. New York: Cambridge University Press (2000).

\bibitem{}  C. G. Timpson, e-print quant-ph/0412063.

\bibitem{}  F. Verstraete and H. Verschelde, Phys. Rev. Lett., {\bf 90}%
(2003)097901.

\bibitem{}  J. D. Zhou, G. Hou, and Y. D. Zhang, Phys. Rev. A, {\bf 64}%
(2001)012301.

\bibitem{}  W. Son, J. H. Lee, M. S. Kim, and Y. J. Park, Phys. Rev. A, {\bf %
64}(2001)064304.

\bibitem{}  J. H. Lee, H. Min, and S. D. Oh, Phys. Rev. A, {\bf 66}%
(2002)052318.

\bibitem{}  A. Grudka, Acta Phys. Slov., {\bf 54}(2004)91.

\bibitem{}  A. Grudka and R. W. Chhajlany, Acta Phys. Pol., {\bf A}\ {\bf 104%
}(2003)409.

\bibitem{}  G. Rigolin, e-print quant-ph/0407219.\ Recently in an e-mail to
the author of this paper Rigolin points out yet that there may be some
correponding relation between $H_{1,2,3,4}^{\left( 16\right) }$\ and $H_{%
{\rm I,II}}^{\left( 16\right) }$\ .

\bibitem{}  W. K. Wootters and W. H. Zurek, Natrue, {\bf 299}(1982)802.
\end{thebibliography}
\end{document}